\documentclass{aa}  
\usepackage[varg]{txfonts}
\usepackage{graphicx}
\usepackage{amssymb,amsmath}
\usepackage{booktabs,multirow}
\usepackage{bm}
\usepackage{natbib}
\bibpunct{(}{)}{;}{a}{}{,}

\usepackage{graphicx}

\begin{document}

\title{Polarity inversion line helicities and solar eruptivity}

\author{K. Moraitis$^1$ \and J.K. Thalmann$^2$}
\institute{Physics Department, University of Ioannina, Ioannina GR-45110, Greece \and University of Graz, Institute of Physics, Universit{\"a}tsplatz 5, 8010 Graz, Austria}

\date{Received ... / Accepted ...}

\abstract{}{This work examines the relationship between solar eruptivity and the relative helicity that is contained around the polarity inversion line (PIL) of the magnetic field, along with its current-carrying component.}{To this end, we analyze the evolution of the PIL helicities in a sample of $\sim 40$ solar active regions which exhibited more than 200 flares of class M or higher. The computation of the PIL helicities is accomplished with the help of relative field line helicity, the recently-developed proxy for the density of relative helicity, following the extrapolation of the 3D coronal magnetic field with a nonlinear force-free method.}{We find that, on average, the relative helicity of the PIL decreases significantly, by more than $10\%$, during stronger eruptive flares (M5.0 class and above), while smaller changes are observed for confined and/or weaker flares. The PIL current-carrying helicity shows higher-magnitude decreases in both strong and weak flares, reaching $20\%$ average changes during the stronger eruptive flares. Notably, the PIL current-carrying helicity displays the most pronounced distinction between eruptive and confined flares, indicating its strong potential as a diagnostic of solar eruptivity. We discuss the implications of these findings for solar flare forecasting.}{}


\titlerunning{PIL helicities and solar eruptivity}
\authorrunning{Moraitis \& Thalmann}

\maketitle

\section{Introduction}
\label{sect:introduction}

Magnetic helicity is a geometrical quantity that describes the complexity of a magnetic field, as it is related to the twist and writhe of individual magnetic field lines, and the intertwining of pairs of field lines. It is conserved in ideal magnetohydrodynamics \citep[MHD;][]{woltjer58}, and therefore important in studies of magnetised systems. The appropriate mathematical form of helicity in solar conditions, where magnetic flux is exchanged between the solar interior and atmosphere, is relative helicity \citep{BergerF84}. This is defined by
\begin{equation}
H_\mathrm{r}=\int_V (\mathbf{A}+\mathbf{A}_\mathrm{p})\cdot (\mathbf{B}-\mathbf{B}_\mathrm{p})\,\mathrm{d}V,
\end{equation}
where $\mathbf{B}$ stands for the 3D magnetic field in the coronal volume of interest $V$, $\mathbf{B}_\mathrm{p}$ for a suitable reference magnetic field, while $\mathbf{A}$ and $\mathbf{A}_\mathrm{p}$ are the respective vector potentials. We refer to $H_\mathrm{r}$ in the following simply as helicity, or volume helicity.

The connection between helicity and solar eruptivity was recognized immediately when helicity began to be used in studies of observed active regions (ARs). It was then found that eruptive flares -those accompanied by coronal mass ejections (CMEs)- have statistically higher helicity than ARs that exhibit confined flares \citep{nindos04}. This work also identified a helicity threshold of $1.5\times 10^{42}\,\rm{Mx}^2$ above which ARs tend to produce CMEs, which was later confirmed by \citet[][$2\times 10^{42}\,\rm{Mx}^2$]{tgr12}, and \citet[][$9\times 10^{41}\,\rm{Mx}^2$]{liokati23}. More recently, \citet{thalmann25} found favorable conditions for large flaring (GOES class M1.0 or larger) when helicity exceeds $1\times 10^{42}\,\rm{Mx}^2$ (see their Fig.~8a), in support of the earlier statistical studies.

A quite successful helicity-derived eruptivity indicator is the so-called helicity eruptivity index \citep[HEI;][]{pariat17}. This is the ratio $|H_\mathrm{j}|/|H_\mathrm{r}|$, where $H_\mathrm{j}=\int_V (\mathbf{A}-\mathbf{A}_\mathrm{p})\cdot (\mathbf{B}-\mathbf{B}_\mathrm{p})\,\mathrm{d}V$ is the current-carrying helicity, one of the two components in the unique decomposition of helicity, the other being the volume-threading helicity \citep{berger99,linan18}. The dedicated numerical experiment by \citet{pariat23} showed that the values of  HEI during the pre-eruptive phase are large for ``eruptive'' cases where the unstable flux rope is later expelled from the simulation volume. In particular, their eruptive simulation setup involved an (overlying) arcade field oriented anti-parallel to the direction of the upper part of the emerging twisted flux rope. In contrast, for simulation setups that involved a quasi-parallel arcade field, the pre-eruption values of the HEI remained low. In line, the simulation-based study by \citet{rice22} showed that the HEI has only a weak predictive skill, and in fact is lower, rather than higher, prior to eruptions when the overlying background magnetic field has the same direction as the arcade. They concluded that, therefore, a whole class of solar eruptions cannot be predicted by a high HEI. Nevertheless, the recent statistical study by \citet{thalmann25}, based on data-constrained coronal magnetic field modelling of 40 solar ARs, revealed that the corona prior to eruptive flares exhibits a larger overall value of the HEI, compared to the characteristic non-eruptive pre-flare corona. Largest values of the HEI, however, were found not just before the onset of CMEs, but persisted over extended time periods so that the exceeding of suggested critical values cannot serve as an indicator for near-in-time upcoming eruptive activity.

With the routine measurements of the magnetic field with the Solar Dynamics Observatory \citep[SDO;][]{pes12} in recent years, the interest in helicity has expanded to include the calculation of the changes it undergoes during solar flares. In a sample of 21 X-class flares, \citet{liu23} found that when a flare is CME-associated it is accompanied by large decreases in helicity ($\sim 17\%$ of the pre-flare values, on average), while this is not the case for confined flares. This was further quantified by \citet{wang23} for 47 flares of GOES class M4.0 or larger, where helicity dropped by $\sim 15\%$ in the CME-associated flares, and by $\sim 1\%$ in the confined. Similar results were obtained by \citet{thalmann25} based on a much more extended flare sample (220 flares of GOES class M1 or larger). Furthermore, for the subset of major flares (GOES class M5 or larger), they found characteristic decreases of helicity of $\sim 12\%$ and $\sim 1\%$, for CME-associated and confined (CME-less) events, respectively.

Two additional helicity-related quantities that have been attracting attention recently are the relative helicity contained around the magnetic polarity inversion line \citep[PIL;][]{moraitis24}, and its corresponding current-carrying component \citep{moraitis24b}. These authors demonstrated that the evolution of the PIL helicities follows in general that of the volume helicities, and that the PIL helicities exhibit a more pronounced change due to the occurrence of flares than the volume helicities. \citet{moraitis24} quantified this difference to an average $\sim 7\%$ decrease for PIL relative helicity and only $\sim 1\%$ for volume helicity, based on their study of 22 CME-associated and CME-less flares above the M1.0 class.

The main aim of this work is to examine the statistical distributions of the changes of the two PIL helicities during solar flares, in the large, and balanced in number of CME-associated and confined flares, flare sample of \citet{thalmann25}, so that to expand on both the works of \citet{moraitis24} and \citet{thalmann25}. In Sect.~\ref{sect:data} we describe the main characteristics of the flare sample that we use. In Sect.~\ref{sect:method} we define all quantities of interest and the methodology for computing them from the observational data. In Sect.~\ref{sect:results} we present the obtained results, and in Sect.~\ref{sect:discussion} we summarize and discuss the results of the paper.

\section{Flare sample}
\label{sect:data}

We study the PIL helicities in the same flare sample as in the work of \citet{thalmann25}. We repeat here its basic characteristics. The analyzed flares are of GOES class M1 or larger, and correspond to 36 different ARs, which are most of the ARs observed with the Helioseismic and Magnetic Imager \citep[HMI;][]{sche12} on board SDO during the rising phase of Solar Cycle 24, from February 2011 to September 2017.

The whole sample consists of 220 M- and X-class flares, 116 of which are confined and 104 CME-associated, and it is denoted as S$_\mathrm{M1+}$. It can be split into two subsets: the weaker flares, and the major flares. The weaker flares are up to the M4.9 class, and are 172 in total, with 102 of them being confined and 70 CME-associated. The major flares (sample S$_\mathrm{major}$) are 48 flares above M5.0, 14 confined and 34 CME-associated. They can also be split into 26 M-class flares (8 confined, 18 CME-associated), and 22 X-class flares (6 confined, 16 CME-associated). The number of flares given here comes from Table 1 of \citet{thalmann25} after subtracting the 11 flares that were associated to a flare-related change of the sign of helicity, and therefore had questionable pre-flare values. More details can be found in \citet{thalmann25}. From all the information of the original flare sample, we only use in our analysis the start and end times of the flares, $t_\mathrm{start}$ and $t_\mathrm{end}$, their strength, and whether a flare is CME-associated or not.

\section{Methodology}
\label{sect:method}

In order to compute the quantities of interest we model the 3D magnetic field in the coronal volume, starting from the observed magnetograms, for each target AR around the time of the flares. This is accomplished with the nonlinear force-free (NLFF) method of \citet{wieg12}, as implemented and described in \citet{thalmann25}. We remind here that the NLFF modelling involves the solution of the force-free boundary value problem on an initially coarse grid with boundaries specified by a potential field solution (computed from the observed vertical magnetic field component). The solution of the coarse grid is then sampled to the final NLFF model grid and serves as a start equilibrium (hence boundary condition). By design, the optimization method weighs the initial equilibrium so that its impact drops in form of a cosine-profile towards the lateral and top boundaries \citep{wieg04}, and therefore, the exact boundary conditions have little, if any, impact on the final NLFF model. We also emphasize that the NLFF reconstruction uses an enhanced weighting for the volume-integrated divergence so that the solutions are of sufficient solenoidal quality, necessary for helicity computations \citep{valori16}. The 3D NLFF models are employed based on a variable time cadence, typically 12~min around flare peak times and 1~hr otherwise. Upon careful check of each of the employed NLFF solutions, unphysical and low quality solutions are identified and excluded from further analysis. 

For all qualifying 3D NLFF models, we compute the reference potential field ($\mathbf{B}_\mathrm{p}$) and the vector potentials of the two fields ($\mathbf{A}$ and $\mathbf{A}_\mathrm{p}$), following the methodology of \citet{moraitis14}. We remind that both vector potentials are taken in the \citet{devore00} gauge, and that $\mathbf{A}_\mathrm{p}$ additionally satisfies the Coulomb gauge. These four vector fields allow us to calculate, among other quantities, the free energy, $E_\mathrm{j}=\int_V (\mathbf{B}-\mathbf{B}_\mathrm{p})^2\,{\rm d}V/(8\pi)$, the volume relative helicity, $H_\mathrm{r}$, and the volume current-carrying helicity, $H_\mathrm{j}$. 

We also compute the relative field line helicity \citep[FLH; $h_\mathrm{r}$;][]{yeates18,moraitis19}, and its respective current-carrying component \citep[$h_\mathrm{j}$;][]{moraitis24b}. These FLHs can be considered as the densities of the respective helicities, and are gauge-dependent in general. The choice of the DeVore gauge facilitates computations, and, as shown in \citet{moraitis24}, produces results to within $10\%$ from other gauge choices. The computation of the FLHs requires to integrate the vector potentials along the magnetic field lines of $\mathbf{B}$ and $\mathbf{B}_\mathrm{p}$. The field line integrations are done with a modification of the \texttt{qfactor} code \citep{liu16}, which can be found online\footnote{\url{https://doi.org/10.5281/zenodo.15689804}}.

The next step is to identify the PIL for each target AR and time of interest. This is done with the method of \citet{schrijver07} and the same parameters as in \citet{moraitis24}, namely, a magnetic field threshold of 150~G, and a dilation window of 3x3 pixels. The convolution of the PIL with an area-normalized Gaussian of 9" full width at half maximum results in a 2D mask $W_\mathrm{PIL}$ at the model lower boundary ($z=0$). With the help of the FLHs we can compute the relative helicity that is contained around the PIL from
\begin{equation}
H_\mathrm{r,PIL}=\int_{z=0} h_\mathrm{r}\,W_\mathrm{PIL}\,{\rm d}\Phi,
\label{flhpil}
\end{equation}
and its respective current-carrying component from
\begin{equation}
H_\mathrm{j,PIL}=\int_{z=0} h_\mathrm{j}\,W_\mathrm{PIL}\,{\rm d}\Phi,
\label{fljpil}
\end{equation}
where ${\rm d}\Phi$ stands for the element of magnetic flux. We refer in the following to these helicities as the PIL relative helicity and the PIL current-carrying helicity, respectively.

Due to the variable time cadence, all quantities of interest are first interpolated to a constant-cadence timeseries of 12~min. Then, the interpolated time arrays are smoothed with a two-hour smoothing window so that sharp changes are eliminated. The effect of these processes can be seen in Fig.~2 of \citet{thalmann25}. Finally, we estimate the flare-related changes of the various quantities. For a quantity $Q$ we define its relative change during a flare as
\begin{equation}
\eta_\mathrm{Q}=\frac{Q_\mathrm{post}-Q_\mathrm{pre}}{Q_\mathrm{pre}},
\end{equation}
where $Q_\mathrm{pre}$ is the mean pre-flare value of $Q$ in the hour prior to the flare start time, i.e., in the interval $t_\mathrm{start}-60\,\mathrm{min}\leq t \leq t_\mathrm{start}$, and similarly, $Q_\mathrm{post}$ is its mean post-flare value in the hour after the flare end time, i.e., in the interval $t_\mathrm{end}\leq t\leq t_\mathrm{end}+60\,\mathrm{min}$. Both $Q_\mathrm{pre}$ and $Q_\mathrm{post}$ are considered with their respective sign. The quantity $\eta_\mathrm{Q}$ is identical with the $\eta$ parameter of \citet{thalmann25} and similar in scope to the relative change $\Delta f$ of \citet{moraitis24}. A negative (positive) value of $\eta_\mathrm{Q}$ denotes a decrease (increase) of $Q$ during that flare.

\section{Results}
\label{sect:results}

With the described methodology we can compute the various quantities, focusing on their flare-related changes, in either of the flare samples of interest. We examine three quantities that are known for their sensitivity with respect to solar eruptivity: free energy, volume relative helicity, and volume current-carrying helicity. These were also examined in \citet{thalmann25} and will be used to interpret our results. Our main interest however is the PIL relative helicity and the PIL current-carrying helicity.

For each of these five quantities we compute the distributions of their flare-related changes ($\eta_\mathrm{Q}$). For each distribution we compute the mean and standard deviation, and also its median value and interquartile range (IQR), that is, the range between the 75th and 25th percentiles. The latter two are insensitive to outlier values that affect the mean and standard deviation, and are thus used as independent measures of the central tendency and dispersion for each distribution. Nevertheless, we limit in the following the level of outlier values to $\pm100\%$, that is, we omit any flare-related changes outside those limits, as we get such extreme values in some cases. This happens when the mean pre-flare values are very small, or when the PIL identification results in very few points.

The mean value for the flare-related change of free energy for the sample of flares S$_\mathrm{major}$ is $\langle\eta_\mathrm{E_\mathrm{j}}\rangle=-10.3\%\pm1.5\%$. When we consider only the CME-associated flares it is $-12.2\%\pm1.8\%$, while it is $-5.7\%\pm2.1\%$ for confined flares. These values are very close to the respective values $-13.3\%\pm5.0\%$ and $-5.2\%\pm2.5\%$ of \citet{thalmann25}. The small differences should be attributed to the different computational methods used in deriving the basic quantities. The median values for free energy are all very close to the respective means, $-11.4\%$, $-13.7\%$, and $-6.5\%$, and so the latter are to be trusted, given also that their IQRs are small, $16.5\%$, $16.3\%$, and $8.8\%$.

\begin{figure*}[ht]
\centering
\includegraphics[width=0.46\textwidth]{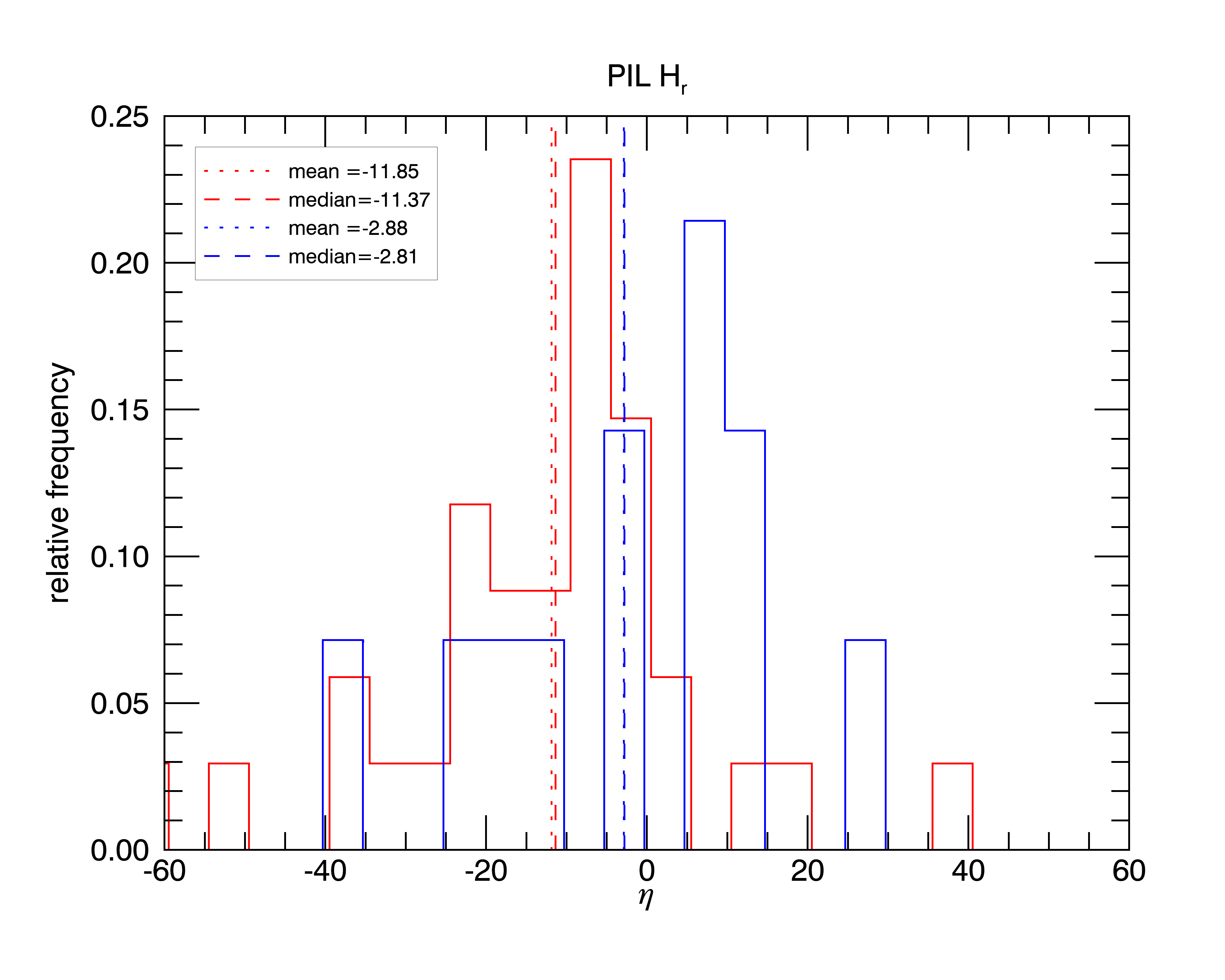}%
\includegraphics[width=0.46\textwidth]{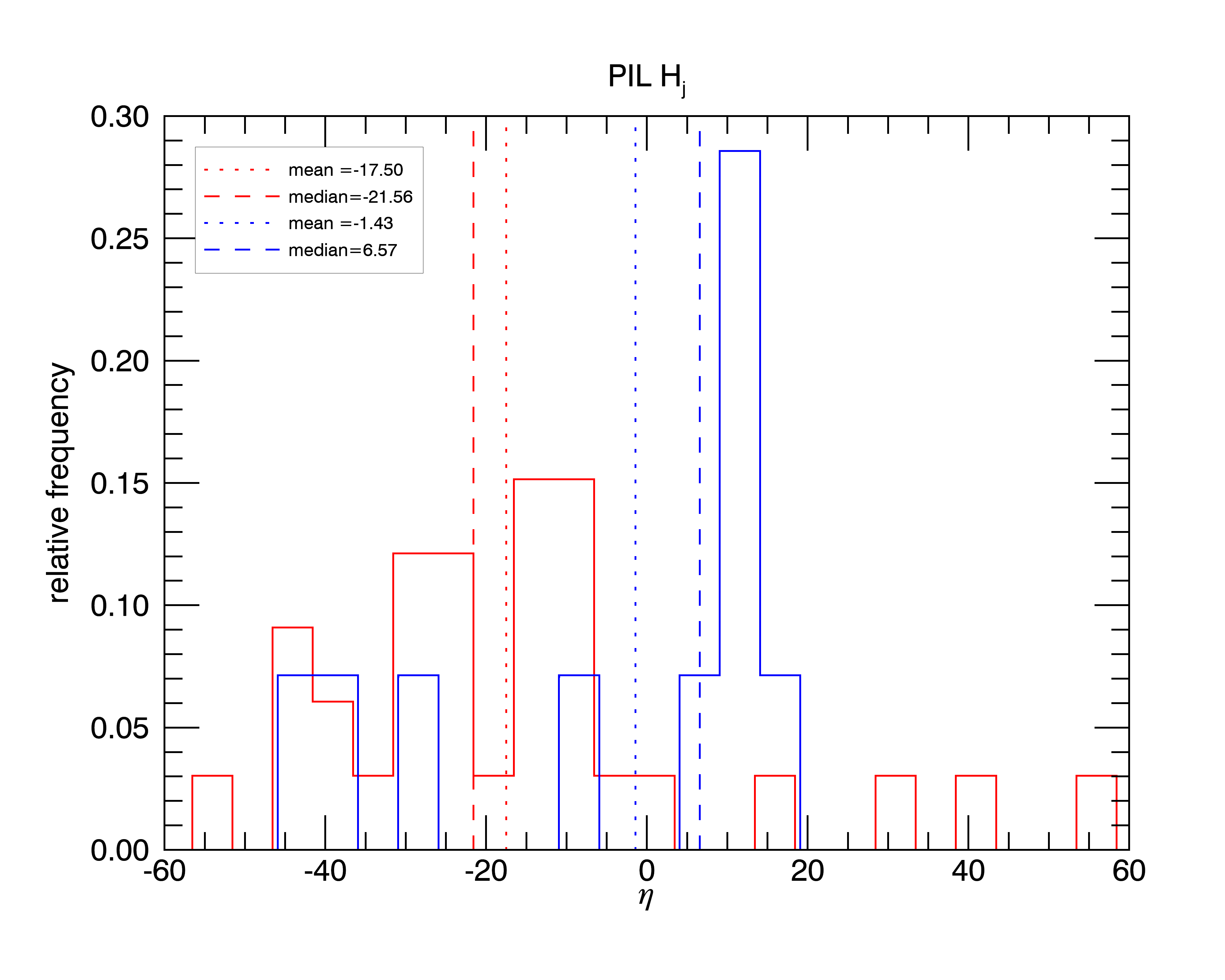}
\caption{Histograms of the relative change during the flares ($\eta$, in \% of pre-flare values), of the PIL relative helicity (left), and of its current-carrying component (right). CME-associated flares are depicted with red and confined ones with blue. The vertical dotted lines represent the mean values of the respective distributions, while the dashed ones the median values.}
\label{histfig}
\end{figure*}

The mean value for the flare-related change of the volume relative helicity is $\langle\eta_\mathrm{H_\mathrm{r}}\rangle=-9.8\%\pm2.1\%$ for the whole S$_\mathrm{major}$ sample, $-13.0\%\pm2.8\%$ for the CME-associated flares, and $-2.3\%\pm1.3\%$ for the confined. The latter two are very close to those of \citet{thalmann25}, $-11.5\%\pm5.0\%$ and $-1.4\%\pm1.9\%$, respectively. As an example of the effect of limiting the outliers to the interval $[-100\%,100\%]$, we mention that the mean $\langle\eta_\mathrm{H_\mathrm{r}}\rangle$ for the CME-associated flares would have been much larger if we had not restricted the outliers, namely $-21.2\%\pm8.6\%$, although the medians would be similar, $-12.0\%$ in the unrestricted case and $-11.8\%$ in the restricted. By applying the limiting however, we get a mean change similar to the median, where the IQR is additionally $16.7\%$. For the confined flares there is disagreement between mean and median values, as the median is $-4.6\%$ with an IQR of $8.1\%$. When we consider all major flares, the median and IQR of the distribution of $\eta_\mathrm{H_\mathrm{r}}$ are $-9.4\%$ and $13.1\%$, respectively.

The volume current-carrying helicity has, overall, strong flare-related changes. For all the major flares it is $\langle\eta_\mathrm{H_\mathrm{j}}\rangle=-16.6\%\pm2.6\%$, for the CME-associated ones $-19.4\%\pm3.3\%$, and for the confined ones $-9.8\%\pm3.5\%$. The respective median values are close to the means, $-14.7\%$, $-21.3\%$, and $-11.3\%$, and the IQRs are larger than in the previous two quantities, namely $23.7\%$, $24.8\%$, and $11.9\%$. The numbers for the CME-associated and confined flares agree very well with those of \citet{thalmann25}, which are $-18.2\%\pm6.5\%$ and $-9.0\%\pm4.1\%$, respectively.

We focus now on the two PIL helicities, starting with the PIL relative helicity, given by Eq.~(\ref{flhpil}). The distribution of the $\eta_\mathrm{H_\mathrm{r,PIL}}$ for all the flares of the sample S$_\mathrm{major}$ is shown on the left panel of Fig.~\ref{histfig}, divided into CME-associated (with red colour), and confined flares (with blue). The mean flare-related change for $H_\mathrm{r,PIL}$ is $\langle \eta_\mathrm{H_\mathrm{r,PIL}}\rangle=-11.8\%\pm3.2\%$ for CME-associated flares, and $-2.9\%\pm10.2\%$ for confined. The median values of the distributions agree with the means, as they are $-11.4\%$ and $-2.8\%$, respectively, while their IQRs are $18.4\%$ and $28.8\%$. If we include all flares, CME-associated and confined, the respective mean $\eta$ value is in between the two, $-9.2\%\pm3.7\%$, with median $-10.6\%$ and IQR $22.7\%$. We notice that the changes of the PIL relative helicity are similar to those of the volume relative helicity in all cases.

The PIL current-carrying helicity, which is given by Eq.~(\ref{fljpil}), is shown on the right panel of Fig.~\ref{histfig}. We find mean flare-related changes $\langle\eta_\mathrm{H_\mathrm{j,PIL}}\rangle=-12.7\%\pm5.3\%$ for all the major flares, $-17.5\%\pm4.7\%$ for the CME-associated ones, and $-1.4\%\pm13.7\%$ for the confined. The respective median values are relatively close to the means for all, and the CME-associated flares, $-17.6\%$ and $-21.6\%$, but differ for the confined flares, where the median is $6.6\%$. Moreover, their IQRs are large, $37.5\%$, $22.4\%$, and $51.9\%$, respectively. The high IQR values indicate that the corresponding distributions are quite broad, that is, they have many outlier values. The comparison of mean and median values indicates that the number of outlier changes of $H_\mathrm{j,PIL}$ is more important for the confined flares. The changes of the PIL current-carrying helicity are similar to those of the volume $H_\mathrm{j}$ for the CME-associated flares ($-17.5\%$ and $-19.4\%$, respectively), but differ a lot for the confined flares ($-1.4\%$ and $-9.8\%$, respectively).

All these results are summarized in Table~\ref{tab1}. For completeness, we give in the lower part of Table~\ref{tab1} the mean and median values of all distributions for the sample S$_\mathrm{M1+}$, that is, for all examined flares. As is expected, all values are lower than their corresponding upper-part ones for the sample S$_\mathrm{major}$. This implies that the changes during the weaker flares are minimal. One can also notice that the highest flare-related decreases are again found in the two current-carrying helicities, volume and PIL, which are both above $7.5\%$.

\begin{table}[ht]
\caption{Characteristics (mean, median, and IQR) of the relative flare-related changes distributions ($\eta_\mathrm{Q}$, in units of \%), divided horizontally into CME-associated, confined, and all flares, and vertically into major and all flares.}
\centering
\resizebox{0.48\textwidth}{!}{
\begin{tabular}{c|ccc|ccc|ccc}
\multicolumn{10}{c}{major flares} \\
\hline
\multirow{2}{*}{$Q$} & \multicolumn{3}{c|}{CME-associated} & \multicolumn{3}{c|}{confined} & \multicolumn{3}{c}{all} \\
\cline{2-10}
 & mean & median & IQR & mean & median & IQR & mean & median & IQR \\
\hline
$E_\mathrm{j}$ & -12.2$\pm$1.8 & -13.7 & 16.3 & -5.7$\pm$2.1 & -6.5 & 8.8 & -10.3$\pm$1.5 & -11.4 & 16.5 \\
$H_\mathrm{r}$ & -13.0$\pm$2.8 & -11.8 & 16.7 & -2.3$\pm$1.3 & -4.6 & 8.1 & -9.8$\pm$2.1 & -9.4 & 13.1 \\
$H_\mathrm{j}$ & -19.4$\pm$3.3 & -21.3 & 24.8 & -9.8$\pm$3.5 & -11.3 & 11.9 & -16.6$\pm$2.6 & -14.7 & 23.7 \\
$H_\mathrm{r,PIL}$ & -11.8$\pm$3.2 & -11.4 & 18.4 & -2.9$\pm$10.2 & -2.8 & 28.8 & -9.2$\pm$3.7 & -10.6 & 22.7 \\
$H_\mathrm{j,PIL}$ & -17.5$\pm$4.7 & -21.6 & 22.4 & -1.4$\pm$13.7 & 6.6 & 51.9 & -12.7$\pm$5.3 & -17.6 & 37.5 \\
\hline
\multicolumn{10}{c}{all flares} \\
\hline
$E_\mathrm{j}$ & -4.8$\pm$1.0 & -6.4 & 12.4 & -2.0$\pm$0.6 & -4.3 & 7.4 & -3.3$\pm$0.6 & -5.0 & 9.3 \\
$H_\mathrm{r}$ & -4.2$\pm$2.0 & -5.8 & 14.2 & -2.4$\pm$1.0 & -3.1 & 7.3 & -3.3$\pm$1.1 & -3.9 & 9.7 \\
$H_\mathrm{j}$ & -7.5$\pm$2.1 & -8.8 & 25.8 & -5.6$\pm$1.5 & -6.1 & 12.8 & -6.5$\pm$1.3 & -7.1 & 17.2 \\
$H_\mathrm{r,PIL}$ & -3.5$\pm$2.6 & -5.6 & 23.9 & -4.0$\pm$1.9 & -4.2 & 15.0 & -3.8$\pm$1.6 & -4.7 & 17.4 \\
$H_\mathrm{j,PIL}$ & -9.1$\pm$3.1 & -12.4 & 29.9 & -5.3$\pm$2.7 & -5.1 & 23.8 & -7.1$\pm$2.0 & -7.8 & 26.4 \\
\hline
\end{tabular}
}
\label{tab1}
\end{table}

We also note that the relative change of $H_\mathrm{r,PIL}$ for all the flares, which has mean value $-3.8\%\pm1.6\%$, median $-4.7\%$, and IQR $17.4\%$, is close to the results of \citet{moraitis24} who find $\Delta f \sim -6.5\%$. The two works are not directly comparable however, due to the different definition of the pre- and post-flare states. These are determined with respect to $t_\mathrm{start}$ and $t_\mathrm{end}$ in this work, while in \citet{moraitis24} the peak of the flares was used to define both pre-, and post-flare states.

Morover, these authors speculated that the relative change of $H_\mathrm{r,PIL}$ would have been larger if only CME-associated flares were considered, and indeed, this is the case in the major flares sample. For the sample S$_\mathrm{M1+}$ however, the differences between CME-associated and confined flares are much smaller.

Finally, we mention that the five examined quantities exhibit flare-related decreases in $\sim 67\%$ of all the CME-associated flares, which increases to $\sim 88\%$ when only the major flares are considered, with the lowest values found in the PIL relative helicity. The respective numbers for the confined flares are much less in all but the free energy and the volume current-carrying helicity, where they are comparable with the CME-associated case.

We also examine the simultaneous decreases of all possible pairs of quantities. The largest value is found for the pair $E_\mathrm{j}$-$H_\mathrm{j}$ in both CME-associated and confined flares, and regardless of whether a flare is a major or a weak one. For the sample S$_\mathrm{M1+}$ we find a simultaneous decrease in $\sim60\%$ of the CME-associated flares and in $\sim57\%$ of the confined flares, while for S$_\mathrm{major}$, these increase to $\sim85\%$ and $\sim79\%$, respectively. These numbers are comparable to those reported by \citet{thalmann25}.

\section{Discussion}
\label{sect:discussion}

This paper examined the statistical distributions of the flare-induced changes of free energy, and of various helicity-related quantities, in a large sample of solar flares. A high-quality NLFF reconstruction of the coronal magnetic field was used in order to compute the relative and current-carrying helicities, both in the entire (flare-relevant) model volume, and associated to the flare-related PIL.

The main results that we got are: 
\begin{itemize}
  \item All examined quantities decrease on average during solar flares. When we differentiate between weaker and major flares, we find that the flare-related changes are much stronger in the latter. As an example, the mean change of $H_\mathrm{r}$ during all flares is $-3.3\%$, while for the major flares it is $-9.8\%$.
  \item The changes resulting from CME-associated flares are always much stronger than in confined flares, especially for major flares.
  \item We confirmed the previous findings by \citet{thalmann25} about the changes of free energy, and of volume relative and current-carrying helicities.
  \item The changes of the PIL relative helicity are comparable to those of the respective volume helicity in all cases, unlike what found in \citet{moraitis24} in a less balanced flare sample. The level of $H_\mathrm{r,PIL}$ changes when we consider all flares however, is similar to that work.
  \item The two current-carrying helicities, volume and PIL, show the most pronounced response to flares. The latter especially, shows the largest differences between CME-associated and confined major flares, $-17.5\%$ compared to $-1.4\%$. Even when the weaker flares are included, the numbers for $H_\mathrm{j,PIL}$ still stand out.
\end{itemize}

The last result can be understood theoretically, as the current-carrying field is mostly related to coronal flux ropes \citep[e.g.,][]{guo17}. In CME-associated flares, where the flux rope is expelled, the current-carrying helicity decreases as it is carried away with the flux rope, while in confined flares it remains largely conserved. This behaviour seems to be captured better by the PIL current-carrying helicity than by the volume one.

Our results reaffirm the importance of the PIL in understanding the creation of solar flares. In addition to the enhanced magnetic flux and electrical currents along the PIL during flares \citep{schrijver07,janvier14}, this region also exhibits a pronounced buildup of current-carrying helicity, which is naturally depicted to the PIL current-carrying helicity. The changes of this quantity thus reflect the changes the coronal field rooted in the PIL experiences during solar flares.

The similar variations in PIL and volume relative helicities may raise the question of whether regions of the AR away from the PIL are just as important as the PIL itself. To examine this possibility, we computed in Appendix~\ref{appx1} the distributions of the flare-related changes of relative and current-carrying helicities that are contained in the region outside of the PIL. The analysis there shows that, although the average flare-related changes of relative helicity inside and outside of the PIL (for the major, CME-associated flares) are comparable, the individual values are much more spread outside of the PIL, and with more values being positive in that case. For the current-carrying helicity the average values are smaller, but still negative, outside of the PIL, and the distribution is, again, much broader than in the PIL case. The bigger spread of these two distributions could be attributed to the much larger range of field line shapes in the region outside of the PIL. For the confined flares, both outside-of-the-PIL helicities exhibit a more compact concentration around 0 compared to the respective PIL cases. We conclude that, although the general trends are similar in the PIL and outside it, regions away from the PIL play a smaller role than the PIL itself, with this effect being more pronounced for current-carrying helicity than for relative helicity.

The PIL helicities have the largest uncertainties of all examined quantities. This is expected because of their computational method, especially the PIL identification algorithm, and it is already noted in \citet{moraitis24b}. Even when we consider the least favourable limits of Table~\ref{tab1} for the PIL current-carrying helicity however, we find that decreases $\gtrsim 13\%$ of this quantity indicate a major CME-associated flare. The criterion $\eta_{H_\mathrm{j,PIL}}\lesssim -13\%$ could thus be used in attempts to forecast imminent space-weather-affecting solar activity. Other factors have to be taken into account of course when considering the possible impact on Earth, such as the AR location on the solar disk, or the relative orientation between the interplanetary and Earth's magnetic fields \citep[e.g.,][]{dumbovic21,thalmann23}. Moreover, a complete study at times without flaring should be carried out as well, so that to eliminate the possibility of `false positives' of the method. Although not addressed in this study, these topics could be examined in a future investigation.

A drawback of using the PIL current-carrying helicity in a forecasting method is the computational effort it requires, which mostly stems from the extrapolation of the magnetic field to the coronal volume \citep[for recent advances see, e.g.,][]{jarolim23}. Potential implications to possibly accelerate the computation of this PIL helicity, will be pursued in future work.

\begin{acknowledgements}
The authors thank the referee for providing constructive comments. KM has received funding from the ERC Whole Sun Synergy grant N$^o$ 810218. JT was funded in part by the Austrian Science Fund (FWF) grants 10.55776/P31413 and 10.55776/PAT7894023. JT acknowledges the High Performance Computing (HPC) center of the University of Graz for providing computational resources and technical support. NASA’s SDO satellite and the HMI instrument were joint efforts by many teams and individuals, whose efforts are greatly appreciated.
\end{acknowledgements}

\bibliographystyle{aa}
\bibliography{refs}

\begin{thebibliography}{31}
\expandafter\ifx\csname natexlab\endcsname\relax\def\natexlab#1{#1}\fi

\bibitem[{{Berger}(1999)}]{berger99}
{Berger}, M.~A. 1999, Plasma Physics and Controlled Fusion, 41, B167

\bibitem[{{Berger} \& {Field}(1984)}]{BergerF84}
{Berger}, M.~A. \& {Field}, G.~B. 1984, J. Fluid. Mech., 147, 133

\bibitem[{{DeVore}(2000)}]{devore00}
{DeVore}, C.~R. 2000, \apj, 539, 944

\bibitem[{{Dumbovi{\'c}} {et~al.}(2021){Dumbovi{\'c}}, {Veronig},
  {Podladchikova}, {Thalmann}, {Chikunova}, {Dissauer}, {Magdaleni{\'c}},
  {Temmer}, {Guo}, \& {Samara}}]{dumbovic21}
{Dumbovi{\'c}}, M., {Veronig}, A.~M., {Podladchikova}, T., {et~al.} 2021, \aap,
  652, A159

\bibitem[{{Guo} {et~al.}(2017){Guo}, {Pariat}, {Valori}, {Anfinogentov},
  {Chen}, {Georgoulis}, {Liu}, {Moraitis}, {Thalmann}, \& {Yang}}]{guo17}
{Guo}, Y., {Pariat}, E., {Valori}, G., {et~al.} 2017, \apj, 840, 40

\bibitem[{{Janvier} {et~al.}(2014){Janvier}, {Aulanier}, {Bommier},
  {Schmieder}, {D{\'e}moulin}, \& {Pariat}}]{janvier14}
{Janvier}, M., {Aulanier}, G., {Bommier}, V., {et~al.} 2014, \apj, 788, 60

\bibitem[{{Jarolim} {et~al.}(2023){Jarolim}, {Thalmann}, {Veronig}, \&
  {Podladchikova}}]{jarolim23}
{Jarolim}, R., {Thalmann}, J.~K., {Veronig}, A.~M., \& {Podladchikova}, T.
  2023, Nature Astronomy, 7, 1171

\bibitem[{{Linan} {et~al.}(2018){Linan}, {Pariat}, {Moraitis}, {Valori}, \&
  {Leake}}]{linan18}
{Linan}, L., {Pariat}, {\'E}., {Moraitis}, K., {Valori}, G., \& {Leake}, J.
  2018, \apj, 865, 52

\bibitem[{{Liokati} {et~al.}(2023){Liokati}, {Nindos}, \&
  {Georgoulis}}]{liokati23}
{Liokati}, E., {Nindos}, A., \& {Georgoulis}, M.~K. 2023, \aap, 672, A38

\bibitem[{{Liu} {et~al.}(2016){Liu}, {Kliem}, {Titov}, {Chen}, {Wang}, {Wang},
  {Liu}, {Xu}, \& {Wiegelmann}}]{liu16}
{Liu}, R., {Kliem}, B., {Titov}, V.~S., {et~al.} 2016, \apj, 818, 148

\bibitem[{{Liu} {et~al.}(2023){Liu}, {Welsch}, {Valori}, {Georgoulis}, {Guo},
  {Pariat}, {Park}, \& {Thalmann}}]{liu23}
{Liu}, Y., {Welsch}, B.~T., {Valori}, G., {et~al.} 2023, \apj, 942, 27

\bibitem[{{Moraitis} {et~al.}(2024{\natexlab{a}}){Moraitis}, {Archontis}, \&
  {Chouliaras}}]{moraitis24b}
{Moraitis}, K., {Archontis}, V., \& {Chouliaras}, G. 2024{\natexlab{a}}, \aap,
  690, A181

\bibitem[{{Moraitis} {et~al.}(2019){Moraitis}, {Pariat}, {Valori}, \&
  {Dalmasse}}]{moraitis19}
{Moraitis}, K., {Pariat}, E., {Valori}, G., \& {Dalmasse}, K. 2019, \aap, 624,
  A51

\bibitem[{{Moraitis} {et~al.}(2024{\natexlab{b}}){Moraitis}, {Patsourakos},
  {Nindos}, {Thalmann}, \& {Pariat}}]{moraitis24}
{Moraitis}, K., {Patsourakos}, S., {Nindos}, A., {Thalmann}, J.~K., \&
  {Pariat}, {\'E}. 2024{\natexlab{b}}, \aap, 683, A87

\bibitem[{{Moraitis} {et~al.}(2014){Moraitis}, {Tziotziou}, {Georgoulis}, \&
  {Archontis}}]{moraitis14}
{Moraitis}, K., {Tziotziou}, K., {Georgoulis}, M.~K., \& {Archontis}, V. 2014,
  \solphys, 289, 4453

\bibitem[{Nindos \& Andrews(2004)}]{nindos04}
Nindos, A. \& Andrews, M.~D. 2004, \apjl, 616, L175

\bibitem[{{Pariat} {et~al.}(2017){Pariat}, {Leake}, {Valori}, {Linton},
  {Zuccarello}, \& {Dalmasse}}]{pariat17}
{Pariat}, E., {Leake}, J.~E., {Valori}, G., {et~al.} 2017, \aap, 601, A125

\bibitem[{{Pariat} {et~al.}(2023){Pariat}, {Wyper}, \& {Linan}}]{pariat23}
{Pariat}, E., {Wyper}, P.~F., \& {Linan}, L. 2023, \aap, 669, A33

\bibitem[{{Pesnell} {et~al.}(2012){Pesnell}, {Thompson}, \&
  {Chamberlin}}]{pes12}
{Pesnell}, W.~D., {Thompson}, B.~J., \& {Chamberlin}, P.~C. 2012, \solphys,
  275, 3

\bibitem[{{Rice} \& {Yeates}(2022)}]{rice22}
{Rice}, O. E.~K. \& {Yeates}, A.~R. 2022, Frontiers in Astronomy and Space
  Sciences, 9, 849135

\bibitem[{{Scherrer} {et~al.}(2012){Scherrer}, {Schou}, {Bush}, {Kosovichev},
  {Bogart}, {Hoeksema}, {Liu}, {Duvall}, {Zhao}, {Title}, {Schrijver},
  {Tarbell}, \& {Tomczyk}}]{sche12}
{Scherrer}, P.~H., {Schou}, J., {Bush}, R.~I., {et~al.} 2012, \solphys, 275,
  207

\bibitem[{{Schrijver}(2007)}]{schrijver07}
{Schrijver}, C.~J. 2007, \apjl, 655, L117

\bibitem[{{Thalmann} {et~al.}(2023){Thalmann}, {Dumbovi{\'c}}, {Dissauer},
  {Podladchikova}, {Chikunova}, {Temmer}, {Dickson}, \& {Veronig}}]{thalmann23}
{Thalmann}, J.~K., {Dumbovi{\'c}}, M., {Dissauer}, K., {et~al.} 2023, \aap,
  669, A72

\bibitem[{{Thalmann} {et~al.}(2025){Thalmann}, {Gupta}, {Veronig}, \&
  {Liu}}]{thalmann25}
{Thalmann}, J.~K., {Gupta}, M., {Veronig}, A.~M., \& {Liu}, Y. 2025, \aap, 695,
  A66

\bibitem[{{Tzio\-tziou} {et~al.}(2012){Tzio\-tziou}, {Georgoulis}, \&
  {Raouafi}}]{tgr12}
{Tzio\-tziou}, K., {Georgoulis}, M.~K., \& {Raouafi}, N.-E. 2012, \apjl, 759,
  L4

\bibitem[{{Valori} {et~al.}(2016){Valori}, {Pariat}, {Anfinogentov}, {Chen},
  {Georgoulis}, {Guo}, {Liu}, {Moraitis}, {Thalmann}, \& {Yang}}]{valori16}
{Valori}, G., {Pariat}, {\'E}., {Anfinogentov}, S., {et~al.} 2016, \ssr, 201,
  147

\bibitem[{Wang {et~al.}(2023)Wang, Zhang, Yang, Yang, \& Zhu}]{wang23}
Wang, Q., Zhang, M., Yang, S., Yang, X., \& Zhu, X. 2023, Res. Astron.
  Astrophys., 23, 095025

\bibitem[{{Wiegelmann}(2004)}]{wieg04}
{Wiegelmann}, T. 2004, \solphys, 219, 87

\bibitem[{{Wiegelmann} {et~al.}(2012){Wiegelmann}, {Thalmann}, {Inhester},
  {Tadesse}, {Sun}, \& {Hoeksema}}]{wieg12}
{Wiegelmann}, T., {Thalmann}, J.~K., {Inhester}, B., {et~al.} 2012, \solphys,
  281, 37

\bibitem[{{Woltjer}(1958)}]{woltjer58}
{Woltjer}, L. 1958, Proceedings of the National Academy of Science, 44, 489

\bibitem[{{Yeates} \& {Page}(2018)}]{yeates18}
{Yeates}, A.~R. \& {Page}, M.~H. 2018, Journal of Plasma Physics, 84, 775840602

\end{thebibliography}

\begin{appendix}

\section{Helicities in the region outside of the PIL}
\label{appx1}

\begin{figure}[ht]
\centering
\includegraphics[width=0.42\textwidth]{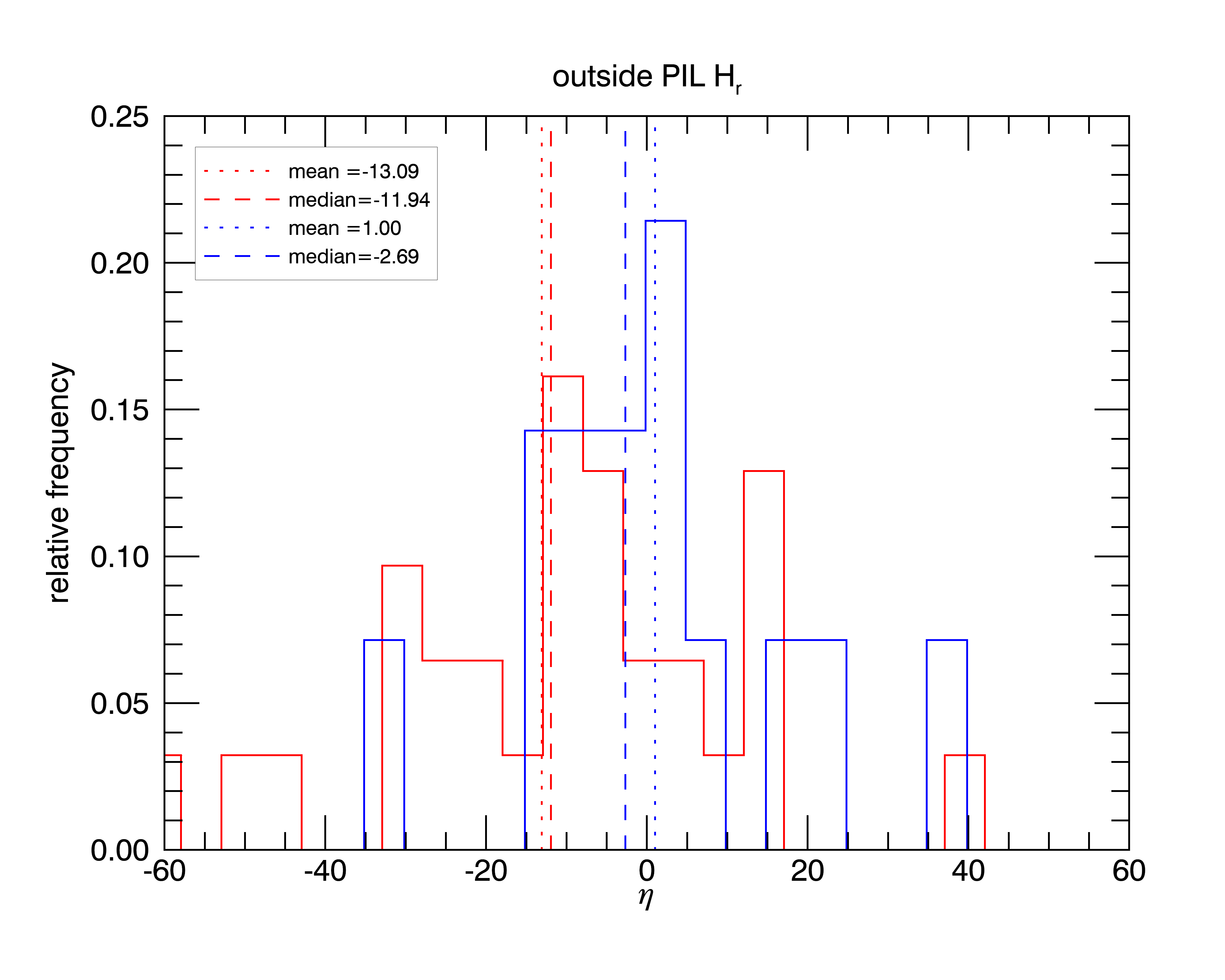}\\
\includegraphics[width=0.42\textwidth]{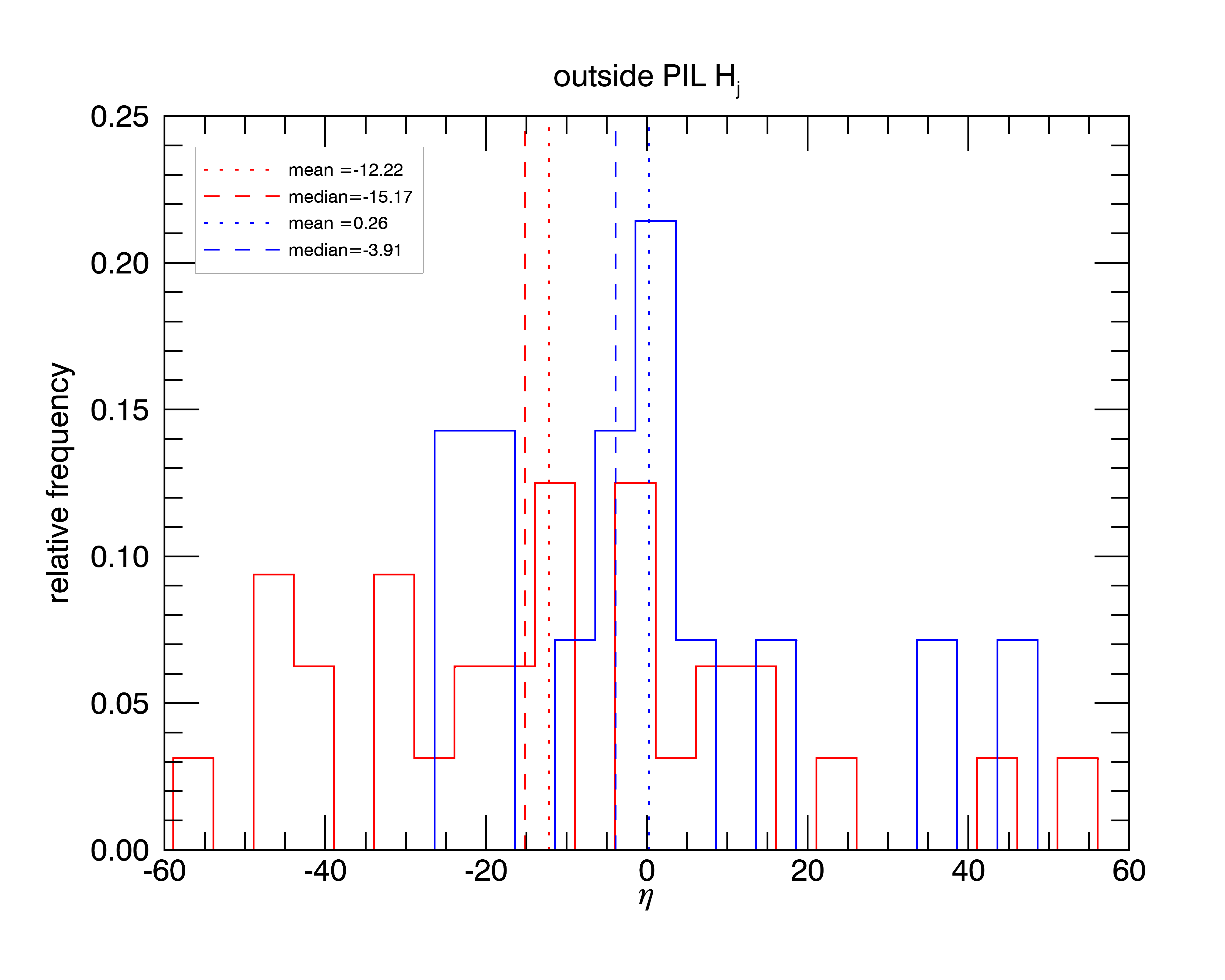}
\caption{Histograms of the relative change during the flares ($\eta$, in \% of pre-flare values), of the relative helicity (top) and of the current-carrying helicity (bottom) that are contained in the region outside of the PIL. CME-associated flares are depicted with red and confined ones with blue. The vertical dotted lines represent the mean values of the respective distributions, while the dashed ones the median values.}
\label{histfig2}
\end{figure}

This Appendix addresses the question of how important is the (relative and current-carrying) helicity content in the region outside of the PIL. We start by defining a complementary to $W_\mathrm{PIL}$ mask, $\bar{W}$, as:
\begin{equation}
\bar{W}=\left\{ 
	\begin{array}{l l}
    0, & \hbox{where } W_\mathrm{PIL}\neq 0 \\
    1, & \hbox{otherwise} \\
	\end{array}
	 \right. .
\end{equation}
This mask selects all field lines on the photosphere other than the PIL. With the help of $\bar{W}$ we can define the relative helicity in the region outside of the PIL from:
\begin{equation}
H_\mathrm{r,out}=\int_{z=0} h_\mathrm{r}\,\bar{W}\,{\rm d}\Phi,
\label{flhpil2}
\end{equation}
and the respective current-carrying helicity, as:
\begin{equation}
H_\mathrm{j,out}=\int_{z=0} h_\mathrm{j}\,\bar{W}\,{\rm d}\Phi.
\label{fljpil2}
\end{equation}
Following the methodology of Sect.~\ref{sect:method} we compute the flare-related changes of these helicities for the major flares, and plot their distributions in Fig.~\ref{histfig2}, separately for CME-associated, and for confined flares.

We immediately notice that the distributions for the CME-associated flares (red curves) are broader compared to the respective ones of Fig.~\ref{histfig}. On the contrary, the distributions for the confined flares (blue curves) seem closer to, and more concentrated around 0 compared to the respective distributions of Fig.~\ref{histfig}.

More specifically, the mean and median values of the $\eta_{H_\mathrm{r,out}}$ distribution for the CME-associated flares are $-13.1\%$ and $-11.9\%$, respectively, not very different from those of $\eta_{H_\mathrm{r,PIL}}$ ($-11.8\%$ and $-11.4\%$). The distribution of $\eta_{H_\mathrm{r,out}}$ is quite broader however, as it has an IQR of $28.8\%$, much larger than the $18.4\%$ of $\eta_{H_\mathrm{r,PIL}}$. For the confined flares, the mean and median values of $\eta_{H_\mathrm{r,out}}$ are $1.0\%$ and $-2.7\%$, and the IQR is $16.3\%$, much less than in $\eta_{H_\mathrm{r,PIL}}$ ($28.8\%$).

The $\eta_{H_\mathrm{j,out}}$ distribution for the CME-associated flares is shifted to the right (closer to 0) with respect to Fig.~\ref{histfig}, as its mean and median values are $-12.2\%$ and $-15.2\%$ ($-17.5\%$ and $-21.6\%$ in Fig.~\ref{histfig}). It is also broader, with an IQR of $38.3\%$ ($22.4\%$ in Fig.~\ref{histfig}). For the confined flares, the $\eta_{H_\mathrm{j,out}}$ distribution is closer to 0 (mean $0.3\%$, median $-3.9\%$), and more concentrated (IQR of $23.8\%$), compared to $\eta_{H_\mathrm{j,PIL}}$ (mean $-1.4\%$, median $6.6\%$, IQR $51.9\%$). All these values are summarized in Table~\ref{tab2} along with the values for $\eta_{H_\mathrm{r,PIL}}$ and $\eta_{H_\mathrm{j,PIL}}$ from Table~\ref{tab1} for comparison.

\begin{table}[ht]
\caption{Characteristics (mean, median, and IQR) of the relative flare-related changes distributions ($\eta_\mathrm{Q}$, in units of \%) for the major flares, divided into CME-associated and confined.}
\centering
\resizebox{0.48\textwidth}{!}{
\begin{tabular}{c|ccc|ccc}
\hline
\multirow{2}{*}{$Q$} & \multicolumn{3}{c|}{CME-associated} & \multicolumn{3}{c|}{confined} \\
\cline{2-7}
 & mean & median & IQR & mean & median & IQR \\
\hline
$H_\mathrm{r,out}$ & -13.1$\pm$4.6 & -11.9 & 28.8 & 1.0$\pm$4.5 & -2.7 & 16.3 \\
$H_\mathrm{r,PIL}$ & -11.8$\pm$3.2 & -11.4 & 18.4 & -2.9$\pm$10.2 & -2.8 & 28.8 \\
$H_\mathrm{j,out}$ & -12.2$\pm$5.7 & -15.2 & 38.3 & 0.3$\pm$5.5 & -3.9 & 23.8 \\
$H_\mathrm{j,PIL}$ & -17.5$\pm$4.7 & -21.6 & 22.4 & -1.4$\pm$13.7 & 6.6 & 51.9 \\
\hline
\end{tabular}
}
\label{tab2}
\end{table}

\end{appendix}

\end{document}